\documentclass[twocolumn,showpacs,preprintnumbers,amsmath,amssymb,aps]{revtex4}
\input epsf
\usepackage{graphicx}
\begin{document}

\title{Vortex configurations and dynamics in elliptical pinning sites
for high matching fields} 
\author{C.J. Olson Reichhardt$^1$, A. Lib{\' a}l$^{1,2}$, 
and C. Reichhardt$^1$} 
\affiliation{ 
$^1$Theoretical Division and Center for Nonlinear Studies,
Los Alamos National Laboratory, Los Alamos, New Mexico 87545\\
$^2$Department of Physics, University of Notre Dame, Notre Dame, Indiana
46556}

\date{\today}
\begin{abstract}
Using numerical simulations
we study the configurations, dynamics, and melting properties of
vortex lattices interacting with elliptical pinning sites at 
integer matching fields with as many as 27 vortices per
pin.  Our pinning model is based on a recently produced
experimental system [G. Karapetrov {\it et al.}, Phys.~Rev.~Lett.~{\bf 95},
167002 (2005)], and the vortex configurations we obtain match well 
with experimental vortex images from the same system.  
We find that the strong pinning sites capture
more than one vortex each, and that the saturation number of
vortices residing in a pin increases with applied field due to the
pressure from the surrounding vortices.  At high
matching fields, the vortices in the interstitial
regions form a disordered triangular lattice.  We 
measure the depinning thresholds for both the $x$ and $y$ directions, and find
distinctive dynamical responses along with highly anisotropic thresholds.
For melting of the vortex configurations under zero applied current, we find
multi-step melting transitions in which the interstitial 
vortices melt at a much lower temperature than the pinned vortices.
We associate this with signatures in the specific heat.
\end{abstract}
\maketitle

\vskip2pc
The immobilization of superconducting vortices at pinning sites is a critical
area of study for device applications, 
since the dissipation that occurs when the vortices move
under an applied current sets a limitation on the effectiveness of 
the superconductors for lossless transport.
Any superconductor contains naturally occurring pinning which may take
the form of nonsuperconducting inclusions, twin boundaries, or other
types of defects.  Over the past few decades, 
there has been a tremendous amount of progress in the fabrication of 
artificial nanoengineered pinning, and particularly lattices of pins.
Due to the possibility of commensuration effects between the vortex
lattice and the pinning lattice, the artificial pinning can be
tremendously more effective
at immobilizing the vortices and enhancing the maximum critical current
than the naturally occurring disorder
in a sample.  
Significant effort has been focused on developing and 
studying different types of periodic pinning, beginning with
one-dimensional pinning created
through thickness modulation of the superconducting film \cite{Daldini}.
This was soon followed by the fabrication of two-dimensional
arrays of holes
\cite{FioryIEEE,Fiory,Metlushko,Baert}.

Large holes can confine more than one flux quantum each
\cite{MoshchalkovPRB,Moshchalkov98}.
Multiple vortices inside a single hole coalesce into a single multiquanta
vortex, since there is no superconducting material to support distinct
cores for each vortex.  In contrast, in blind holes, 
which do not pass completely through the superconducting
material,  individual vortex cores retain their identity
within the hole.  Multiple occupation by single quantum vortices
in large blind holes was imaged experimentally using a Bitter decoration
technique \cite{Bezryadin} and later studied in simulation
\cite{reichhardt00}.

Continuing advances in nanolithography focused attention on the
production of small, closely spaced holes that could each capture
a single vortex.  Combined with this,
significant advances in imaging techniques have contributed
to our understanding of the interaction between vortices and the periodic
pinning.  
In particular, if the applied magnetic field exceeds the matching field
at which the number of vortices present equals the number of pinning sites,
then in the case of small holes, the extra vortices above the matching
field cannot occupy the holes but must instead sit at interstitial
locations in the superconducting material surrounding the holes. 
These interstitial vortices are weakly pinned via the repulsion from
vortices trapped at pinning sites, and their configurations have been
imaged using
electron-beam microscopy techniques \cite{Harada}, 
scanning Hall probe microscopy \cite{Grigorenko,VanBael01,Field} and
scanning $\mu$SQUID devices \cite{Veauvy}.
The interstitial vortex arrangements are sensitive to the structure of the
periodic pinning, and
a wide range of different types of artificial pinning lattices
have been tested experimentally \cite{Lin96,Morgan,VanBael2,Welp02},  
including both symmetric pinning arrays 
as well as asymmetric arrays such as rectangular
arrangements of holes \cite{Martin99,Stoll}.  The weak pinning of the
interstitial vortices can produce
asymmetric critical currents 
\cite{reichhardt01c,Velez02,Pannetier03,Welling}.

\begin{figure*}
\includegraphics[width=5.0in]{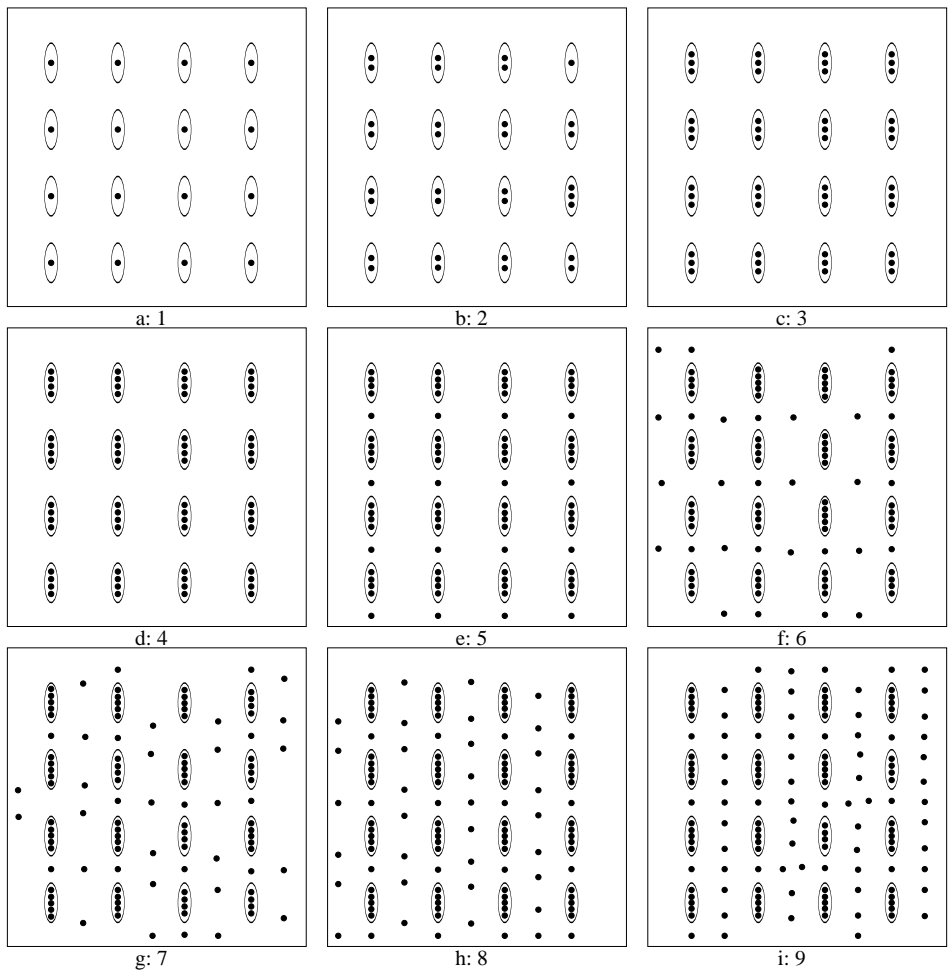}
\caption{
Vortex configurations at
the first nine matching fields.  Black dots indicate vortex positions and
pinning sites are indicated by ellipses.  The pinning troughs connecting
the pins in the $y$ direction are not directly indicated in the figure but
can be identified based on the vortex positions.
$B/B_\phi=$ (a) 1, (b) 2, (c) 3, (d) 4, (e) 5, (f) 6, (g) 7, (h) 8,
and (i) 9.
\label{imagefig}
}
\end{figure*}

Simulations of vortices in periodic arrays of singly occupied pins give
good agreement with experimental observations and indicate the
importance of the interstitial vortices in determining
the dynamical response of the system 
\cite{paps,reichhardt97,ourharada,reichhardt98}.
This is particularly important in understanding the field dependence
of the behavior of nanopatterned superconductors.  Vortices are
strongly pinned at the matching field when the number of vortices
equals the number of artificial pins, but the effectiveness of the pinning
is dramatically reduced whenever defects in the form of interstitials or
vacancies are introduced by shifting the applied field away from its
matching value.  The result is experimentally observable as oscillations
in critical current \cite{Castellanos} or resistivity
\cite{Martin,Jaccard,Martin00}, as well as through separate depinning 
transitions for interstitial vortices
\cite{Rosseel,Metlushko98,Metlushko99,VanLook99}.
Due to the fact that the pinning of interstitial vortices is weak,
pins which capture multiple vortices are much more effective at increasing the
critical current over a wider range of field \cite{reichhardt01}.
There is, however, a trade off between saturation number of the pin and size of
the nanoarrays; for example, in an imaging experiment for relatively
large dots, as many as 280 vortices were captured by each artificial
pin \cite{Surdeanu}.

In the interest of combining small size with greater pinning effectiveness,
recent work has shifted away from simple circular holes or antidots
to asymmetric pinning sites.
Asymmetry can be introduced by placing symmetric pins in 
asymmetric unit cells containing two pins each 
\cite{Zhu01,Vandevondel};  alternatively, the pins themselves may be
made into asymmetric shapes, such as triangles
\cite{Villegas,ourtriangle}
or rectangles \cite{VanLook}. 
Arrangements of rectangular
pins are of particular interest since they may be used as the basis
for vortex logic devices, in which each elongated pin can hold a single
bit of information \cite{RCA,RCA2}.

A new fabrication technique has recently been developed to create well
controlled elongated pinning sites in a sample with an atomically
flat surface \cite{Karapetrov,KarapetrovAPL}.  
This permits direct imaging of
the vortex positions using scanning spectroscopy methods.
Due to the extremely high
resolution of this imaging technique, 
vortex configurations can be observed at fields well above matching, 
showing interesting effects such as the formation of
a disordered triangular vortex lattice in the region between
pinning sites.  Obtaining an image with a scanning technique
is a time-consuming process, so dynamical information about the
vortices is not available from the experiments
of Refs.~\cite{Karapetrov,KarapetrovAPL}.  It would be 
interesting to explore the dynamical behavior of the vortices
under applied currents or finite temperatures at 
high matching fields.
Existing simulations have been confined to low matching fields,
with studies reaching at most the ninth matching field \cite{ourharada},  
and higher matching fields have never been considered.  The
particular pinning geometry used in Ref.~\cite{Karapetrov} also
has unique aspects which have not been explored previously.

\begin{figure*}
\includegraphics[width=5.0in]{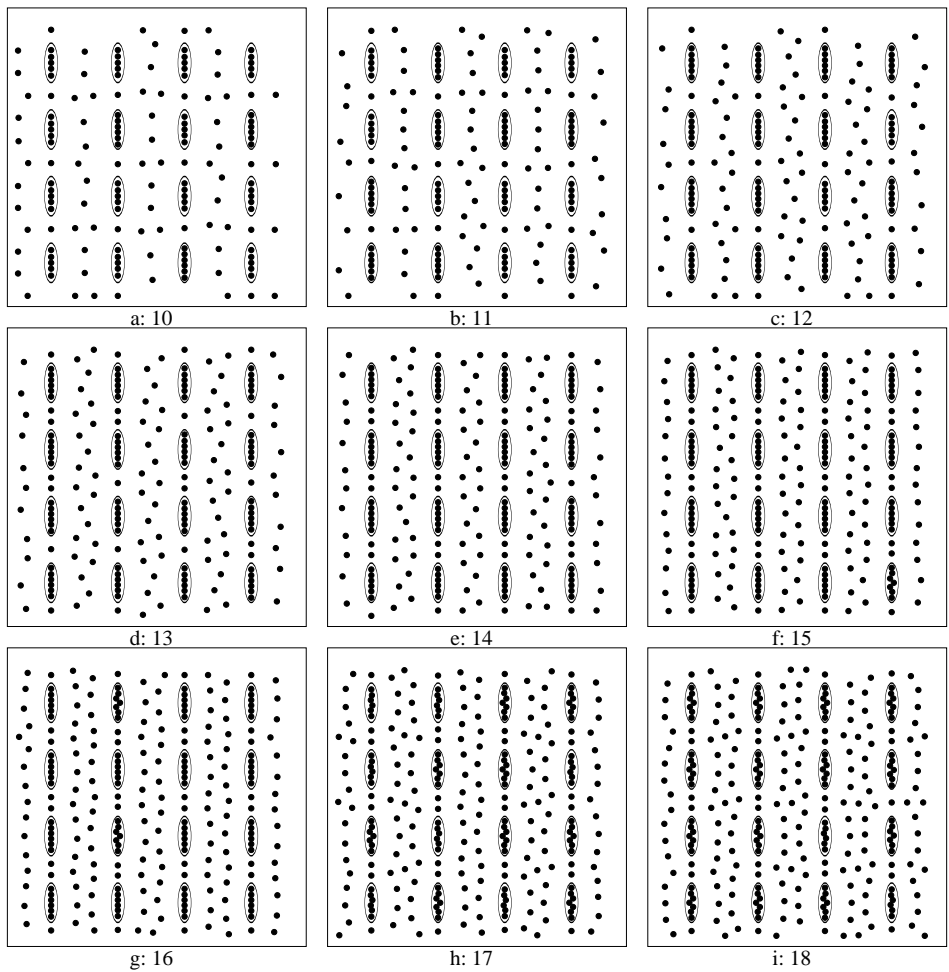}
\caption{
Vortex configurations at 
the second nine matching fields.  Black dots indicate vortex positions and
pinning sites are indicated by ellipses.
$B/B_{\phi}=$ (a) 10, (b) 11, (c) 12, (d) 13, (e) 14, (f) 15,
(g) 16, (h) 17, and (i) 18.
}
\label{imagefig2}
\end{figure*}

In this work, we study the static and dynamic behaviors of vortices
interacting with elongated pinning sites of the type created in
Ref.~\cite{Karapetrov} for matching fields as high as 27 vortices
per pinning site.  We explore the dynamic response of the
vortices to applied currents, and find strongly asymmetric
depinning forces.  We also study the melting of the vortex configurations
and find interesting two-stage melting behavior in which the interstitial
vortices melt at much lower temperatures than the pinned vortices.
The particular model of the pinning that we use has not been
considered before and gives good agreement with the vortex configurations
obtained in experiment.  We also study much higher matching fields
than have been simulated previously.  The dynamic vortex responses that
we observe are related to the particular symmetries present at each
matching field.

\begin{figure*}
\includegraphics[width=5.0in]{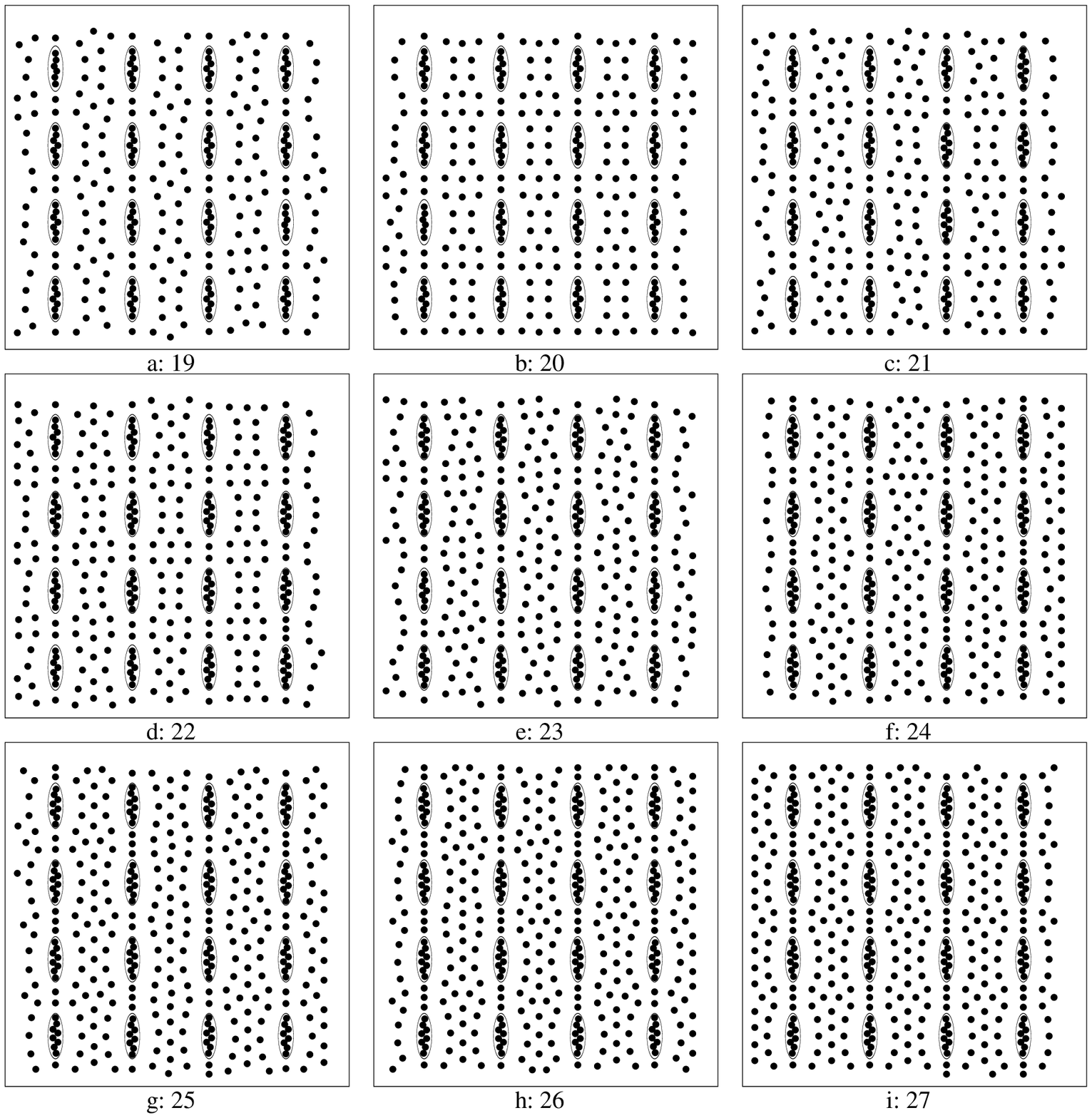}
\caption{
Vortex configurations at
the third nine matching fields.  Black dots indicate vortex positions and
pinning sites are indicated by ellipses.
$B/B_\phi=$ (a) 19, (b) 20, (c) 21, (d) 22, (e) 23, (f) 24, (g) 25,
(h) 26, and (i) 27.
}
\label{imagefig3}
\end{figure*}

We model a two-dimensional system of 
$N_v$ vortices in a superconducting crystal containing $N_p$ artificial
pinning sites.
We assume periodic boundary conditions in the $x$ and $y$ directions.
The  
overdamped equation of motion for an individual vortex $i$ is 
\begin{equation}
\eta\frac{ d{\bf r}_{i}}{dt} = {\bf f}_{i}^{vv} + {\bf f}^{T}_{i} + {\bf f}^{p}_{i} + {\bf f}^{d}_{i} 
\end{equation}
The damping constant $\eta=\phi_0^2d/2\pi\xi^2\rho_N$ in a
crystal of thickness $d$.
Here, $\phi_0=h/2e$ is the flux quantum, $\xi$ is the superconducting
coherence length, and $\rho_N$ is the normal state resistivity of the
material.
The vortex-vortex interaction force is 
\begin{equation}
{\bf f}_{i}^{vv} = \sum^{N_{v}}_{j\ne i}f_{0}
K_{1}\left(\frac{r_{ij}}{\lambda}\right){\bf {\hat r_{ij}}} , 
\end{equation}
where $K_{1}$ is the modified Bessel function appropriate for stiff
three-dimensional vortex lines.
$\lambda$ is the London penetration depth, 
$f_{0}=\phi_0^2/(2\pi\mu_0\lambda^3)$,
and $r_{ij}$ is the distance between
vortices $i$ and $j$.
We measure time in units of $\tau=\eta /f_0$.
For a NbSe$_2$ crystal 0.1 mm thick, we take $\eta=2.36\times 10^{-11}$ Ns/m, 
$f_{0}=6.78\times 10^{-5} N/m$, and
$\tau=0.35$ $\mu$s.
The thermal force ${\bf f}^{T}_{i}$ arises from random 
Langevin kicks with the properties 
$\langle{\bf f}^{T}_{i}\rangle = 0$ and
$\langle{\bf f}_{i}^{T}(t){\bf f}_{j}^{T}(t^{\prime})\rangle 
= 2\eta k_{B}T \delta(t-t^{\prime})\delta_{ij}$. 
The conversion from physical temperature to simulation units is
given by
\begin{equation}
f_{T}=\frac{1}{d}\sqrt{\frac{2\eta k_B T}{\delta \tau}}.
\end{equation}
Using $\delta \tau=0.01$, a temperature of 4.2K corresponds
to $f_{T}=4.12\times 10^{-7} f_0$.  Thus, for our measurements of transport
characteristics, we take $f_{T}=0.$

The quenched disorder ${\bf f}_{i}^{p}$
is modeled as composite pinning sites constructed to match the experimental
geometry.  The sample contains parabolic troughs $0.5\lambda$ wide that
run in the $y$ direction and are spaced 2.5$\lambda$ apart
in the $x$ direction, corresponding
to 0.5$\mu$m in NbSe$_2$.  Each trough is decorated with parabolic traps
that confine only in the $y$ direction, which define the locations of the
pinning sites shown in the experimental images.  These traps are 
$1.5\lambda$ long in the $y$ direction and equal to the trough width
in the $x$ direction.  
The long troughs represent the fact that the crystal layer 
in Ref.~\cite{Karapetrov} is extremely
thin between the pins in the $y$ direction.  This produces an 
energetically favorable
location for a vortex.  Within the
pinning sites, the order parameter is completely suppressed by the
gold in the actual experimental sample, and the vortices inside the pinning
site would form a macrovortex state which is, however, not
circular due to the pinning geometry.  In our model, macrovortex formation
is not permitted, but the configuration of vortices inside a given pin
approximates the noncircular current configuration expected to occur in
the experimental geometry.
We use pinning strengths of 12.0 for the decorating troughs and
90.0 for the channel.
This produces a square array of pinning sites each of which is elongated in
the $y$ direction, as illustrated in Fig.~\ref{imagefig}.

The Lorentz driving force from an external applied current is 
given by ${\bf f}^{d}$, which may be applied either in the $x$ direction,
${\bf f}^d=f_d{\bf {\hat x}}$, or in the $y$ direction, 
${\bf f}^d=f_d{\bf {\hat y}}$.  We measure the corresponding
velocity responses in either $x$,
$V_x=(1/N_v)\langle\sum_{i=1}^{N_v}{\bf v}_{i}\cdot {\bf {\hat x}}\rangle$,
or $y$,
$V_y=(1/N_v)\langle\sum_{i=1}^{N_v}{\bf v}_{i}\cdot {\bf {\hat y}}\rangle$.
The system size is measured in units
of $\lambda$ and the forces in terms of $f_{0}$.
Most of the results presented here are for a sample of size 
$10\lambda \times 10\lambda$
containing $N_p=16$ pins.  We consider filling fractions up to
$B/B_\phi=27$, corresponding to 432 vortices.
At these fields, even though
the vortex-vortex interaction is of a short-range form appropriate
for a crystalline sample, many vortices fall within the interaction
range of a given vortex, and as a result
the simulations become computationally intense.
It is for this reason that we consider the sample size illustrated in
the figures.
We have also tested larger samples of
size $30\lambda \times 30\lambda$ containing 144 pinning sites
and find substantially identical results.

\begin{figure}
\includegraphics[width=3.5in]{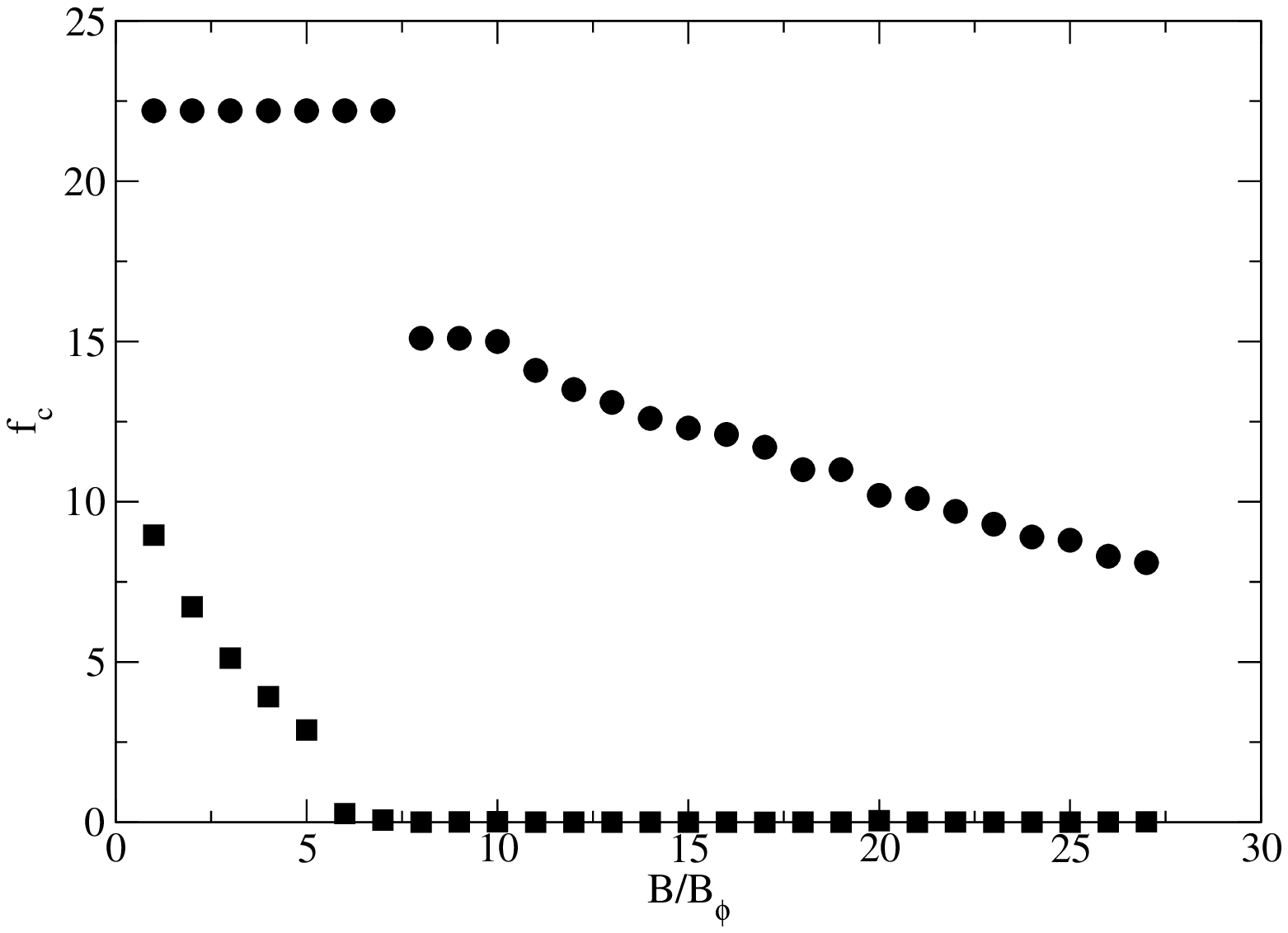}
\caption{
The critical depinning force for two different driving directions
at matching fields $B/B_\phi=1$ to 27.  
Circles: $f_c^x$ for driving in the $x$ direction; squares: $f_c^y$
for driving in the $y$ direction.
}
\label{depin}
\end{figure}

We first consider the vortex arrangements at integer filling fractions
$B/B_\phi=N_v/N_p$.  
For each field, we initialize the sample at a high temperature so that the
vortices are diffusing freely, and then slowly decrease the temperature
in order to anneal the system toward the $T=0$ ground state.
Figure~\ref{imagefig} illustrates the zero temperature
vortex configurations at the first nine matching
fields for a sample with $N_p=16$ pinning sites.
For $B/B_\phi=1$ to 4, the vortices are confined fully within
the pinning sites and form single lines.  At 
$B/B_\phi=5$,
the pinning confinement is not strong enough to capture a fifth vortex
inside the pin, so one vortex per unit cell occupies an interstitial
position.  If the pinning force existed only within the ellipses drawn
in the figure, the interstitial vortex would occupy the low energy point
at the same $y$ position located halfway between rows of pins, but it would 
move
over by half a pinning lattice constant in order to sit midway between
columns of pinning.  This does not happen due to the 
presence of the additional pinning trough
introduced in our model.  A similar vortex configuration for one 
interstitial vortex per unit pinning cell
is observed in the
experiment due to the nonuniform ion beam patterning of the pinning
geometry in the $x$ and $y$ directions, and the configuration in
Fig.~\ref{imagefig}(e) matches well with the configuration shown in
Fig.~2(b) of Ref.~\cite{Karapetrov}.  At 
$B/B_\phi=6$,
a second interstitial vortex can not fit inside the pinning trough, so
it occupies the next most favorable
location centered between four pinning sites.  
The result is the formation of rows of interstitial vortices, and
Fig.~\ref{imagefig}(f) resembles the configuration shown
in Fig.~2(c) of Ref.~\cite{Karapetrov}.  
At 
$B/B_\phi \ge 6$,
we begin to observe some disorder in our vortex configurations, and
for $B/B_\phi=6$, a few
of the pins have captured five vortices instead of four.  This 
trend continues at higher fillings: the occupation number of the pins
does not saturate but slowly increases as the overall higher
vortex density outside the pinning site helps to stabilize a larger number
of vortices inside the pinning site.  
A gradual increase in hole saturation number was also observed in
the experiments of
Ref.~\cite{Karapetrov}. 
At 
$B/B_\phi=7$ the
interstitial vortices begin to spread out into the free channel between
columns of pinning sites, and this continues at 
$B/B_\phi=8$, where all pins now capture five vortices.  
The interstitial vortex channels in
Fig.~\ref{imagefig}(h) match well with Fig.~2(d) of Ref.~\cite{Karapetrov}.
By 
$B/B_\phi=9$, 
some defects occur in the interstitial channels and the vortices start to
buckle out along the space between pinning rows in order to form a
second interstitial column.
Figures \ref{imagefig}(i) is in good agreement with the configuration
obtained in Fig.~3(a) of Ref.~\cite{Karapetrov}.

\begin{figure}
\includegraphics[width=3.5in]{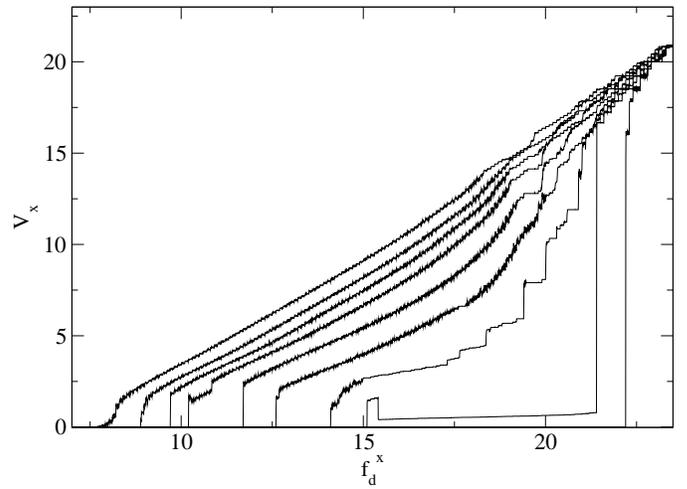}
\caption{
Representative velocity-force curves for x-direction driving, $V_x$
versus $f_d^x$, for $B/B_\phi=$ (from right to left) 1, 9, 11, 14,
17, 20, 22, 24, and 27.
}
\label{fig:iv}
\end{figure}

Vortex configurations at
the second nine matching fields, $B/B_\phi=10$ to 18, 
are illustrated in
Fig.~\ref{imagefig2}.  
For $B/B_\phi=10$ and 11, we find a gradual buckling transition of the
columns of interstitial 
vortices filling the vertical gaps between the pins.  The vortex 
configuration in the columns
gradually changes from one vortex wide at $B/B_\phi=10$ in
Fig.~\ref{imagefig2}(a)
to two vortices wide by $B/B_\phi=15$ in Fig.~\ref{imagefig2}(f).
At $B/B_\phi=16$ in Fig.~\ref{imagefig2}(g) we obtain a configuration
that agrees well with Fig.~3(b) of
Ref.~\cite{Karapetrov}.
The buckling transition begins anew as the field is further increased,
and by $B/B_\phi=18$ in Fig.~\ref{imagefig2}(i)
portions of the interstitial 
channels contain a vortex arrangement three vortices wide.
The pinning sites continue to capture larger numbers of vortices as
the field increases, and at 
$B/B_\phi=15$ and above these
configurations also begin to buckle inside the pinning sites.  The small
portion of the pinning channel between the pins in the $y$ direction also
captures more than one vortex starting at 
$B/B_\phi=13$.

We illustrate vortex configurations at
matching fields much higher than have been simulated
previously, through $B/B_\phi=27$, 
in Fig.~\ref{imagefig3}.  
Vortices continue to populate the interstitial region of the
channels as the field is further increased, and the vortex occupation
number of the pins and of the small trough region connecting the pins
continues to slowly increase with $B/B_\phi$.  We do not find perfectly
parallel rows of vortices as was indicated schematically in Fig.~3(c)
of Ref.~\cite{Karapetrov}, but instead observe a marked modulation of
the vortex density inside the channel caused by the strong repulsion
from the large number of vortices confined within each elongated
pin.  This is in general agreement with the actual image in Fig.~3(c)
of Ref.~\cite{Karapetrov}, which compares well with 
our Fig.~\ref{imagefig3}(e) at $B/B_\phi=23$.
There is still a tendency for the 
interstitial vortices to form rows parallel with
the $x$ axis and passing between pinning rows,
which is particularly pronounced at $B/B_\phi=20$ in
Fig.~\ref{imagefig3}(b) and to a lesser degree at $B/B_\phi=22$ in
Fig.~\ref{imagefig3}(c).  As we will show below, this row structure
affects the depinning properties of the vortex lattice.

\begin{figure}
\includegraphics[width=3.5in]{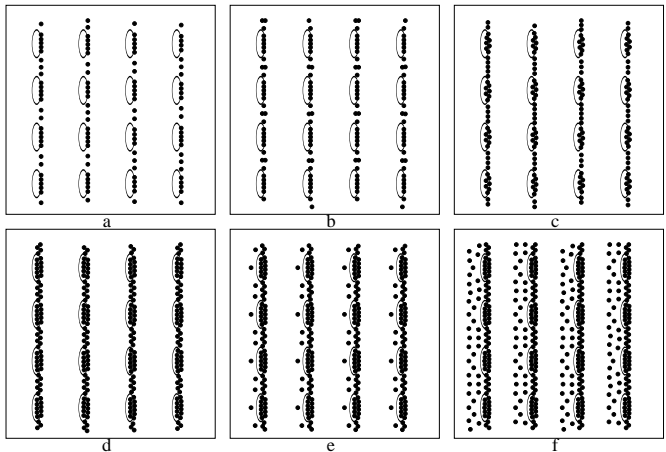}
\caption{
Illustration of the vortex configurations just before depinning under
application of a driving force in the $x$ direction
(to the right in the figure).  Black dots indicate
vortex positions and pinning sites are indicated by ellipses; the pinning
troughs between the pins running 
in the $y$ direction are not directly indicated in the figure.
$B/B_\phi=$ (a) 7, (b) 9, (c) 11, (d) 17, (e) 20, and (f) 27.
}
\label{fig:fcxpix}
\end{figure}

We next study the transport properties of the different fillings by
applying a transport current in either the $x$ or $y$ direction.
Due to the strong anisotropy of the pinning geometry, we find
strongly anisotropic depinning thresholds $f_c^x$ in
the $x$ direction and $f_c^y$ in the $y$ direction.
Figure \ref{depin} illustrates the depinning force at the matching fields
for ${\bf f}_{d}=f_{d}{\bf {\hat x}}$ and ${\bf f}_{d}=f_{d}{\bf {\hat y}}$.
The depinning force in the $y$ direction
$f_c^y$ is much smaller than $f_c^x$ at all matching fields, 
and $f_c^y$ drops nearly to zero above $B/B_\phi=5$ once
interstitial vortices begin to appear in the free channel region between the
pins.  At incommensurate fields, not shown, we expect the depinning force
for both $x$ and $y$ directions
to be lower than it is at the commensurate fields.

\begin{figure}
\includegraphics[width=3.5in]{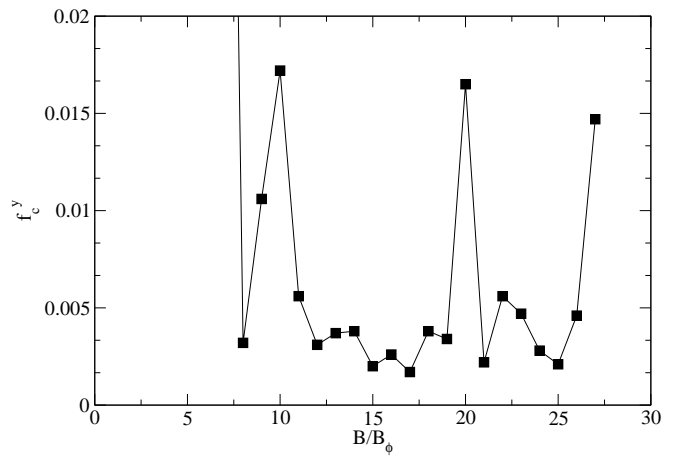}
\caption{
Critical depinning force $f_c^y$ at higher fillings
$B/B_\phi \ge 8$.  The $y$ axis of the figure has been magnified
relative to Fig.~4.
}
\label{fig:depiny}
\end{figure}

We find a nearly constant $f_c^x$ for $B/B_\phi \le 7$, while
for $B/B_\phi \ge 8$, $f_c^x$ gradually decreases with $B/B_\phi$.
Representative velocity-force curves for $x$ direction driving are
illustrated in Fig.~\ref{fig:iv}.  The curves for $B/B_\phi=$ 2 through
7 are essentially identical to the curve shown for $B/B_\phi=1$, with a
sharp depinning transition 
at $f_d^x=22.2$ followed by an ohmic response.
The transition in $f_c^x$ at $B/B_\phi \ge 8$ 
occurs due to a change in the depinning
mechanism.  For $B/B_\phi \le 7$, vortices initially accumulate inside
the pins and the pinning troughs upon application of a driving force, 
as illustrated in Fig.~\ref{fig:fcxpix}(a) for $B/B_\phi=7$.
Depinning occurs when the pinning force of the pinning troughs in the
$x$ direction is overcome.  When $B/B_\phi \ge 8$, not all of the
vortices can fit in a single column inside the trough because the confining
force of the pinning troughs
is insufficient to overcome the vortex-vortex repulsion.  Instead,
the vortex column buckles in the trough area between pins, as illustrated
in Fig.~\ref{fig:fcxpix}(b) for $B/B_\phi=9$.  Depinning initiates at the
buckled points, where the vortex-vortex repulsive force is now added
to the driving force on a buckled vortex; as a result, 
the depinning threshold is
considerably lower than for $B/B_\phi < 8$
and decreases with increasing field.  At higher
fields the buckling in the vortex configurations below depinning
shifts from outside the pins to inside the pins, as
shown in Fig.~\ref{fig:fcxpix}(c) for $B/B_\phi=11$, and eventually
buckling occurs in both locations, as 
illustrated in Fig.~\ref{fig:fcxpix}(d) for
$B/B_\phi=17$.  
At higher fields, all of the
vortices are no longer able to fit inside the pinning trough before the
depinning threshold is reached, and instead the interstitial vortices
pile up against the repulsive barrier formed by the vortices inside the pinning
trough, as shown in Fig.~\ref{fig:fcxpix}(e) and (f) for $B/B_\phi=$ 20
and 27.
In this regime, the pinning resembles extremely strong twinning barriers.

At $B/B_\phi=9$, close to the transition where the initial buckling 
of the vortex configuration before depinning shifts
from the location between the pins to within the pins, we observe a negative
$dV/dI$ characteristic as indicated in Fig.~\ref{fig:iv}
around $f_d^x=16$.  Here, after depinning the vortices
initially flow between the rows of pinning sites at the location of the
buckled points, but as the drive increases, the flow location shifts and
the moving vortices instead pass directly 
through the pinning sites while the vortices 
in the troughs between the pinning sites remain pinned.  At higher drives
there is a step up in $V_x$ at the point when all of the vortices depin.
This feature is similar to the dynamic phases observed for symmetric
periodic pins in  
Ref.~\cite{reichhardt97}.  For fields above $B/B_\phi=9$, 
depinning initiates at the pinning sites and not in the troughs between
pins.
The initial
vortex motion passes through the pinning sites, and the vortices in the
troughs between pins do not depin until higher drives are applied.
At fields of $B/B_\phi=17$ and higher, when interstitial vortices
are always present even just below the depinning transition, the depinning 
gradually becomes more elastic with increasing field, as indicated by the
gradual rounding of the velocity force curves at depinning in Fig.~5.

\begin{figure}
\includegraphics[width=3.5in]{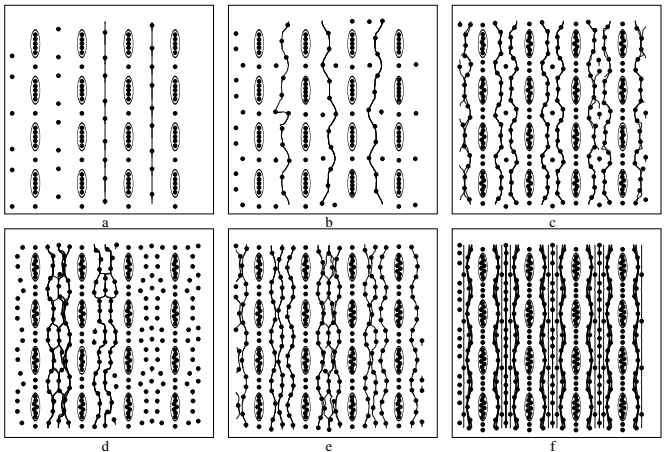}
\caption{
Illustration of the vortex motion just after depinning under
application of a driving force in the $y$ direction
(to the top in the figure).  Black dots indicate
vortex positions, pinning sites are indicated by ellipses, and
lines indicate trajectories followed by the vortices over a
period of time.
$B/B_\phi=$ (a) 8, (b) 10, (c) 19, (d) 20, (e) 21, and (f) 27.
}
\label{fig:fcypix}
\end{figure}

\begin{figure}
\includegraphics[width=3.5in]{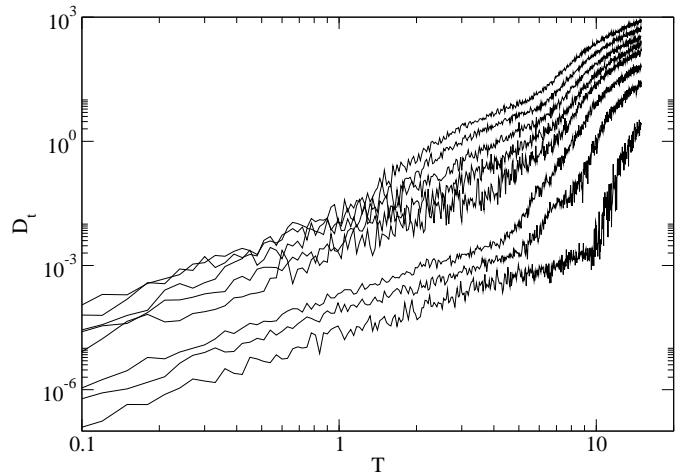}
\caption{
Finite time total diffusion $D_t$ as a function of temperature 
for representative vortex fillings $B/B_\phi=$ (bottom to top) 
1, 3, 5, 8, 10, 13, 17, and 22.
}
\label{diffuse}
\end{figure}

\begin{figure}
\includegraphics[width=3.5in]{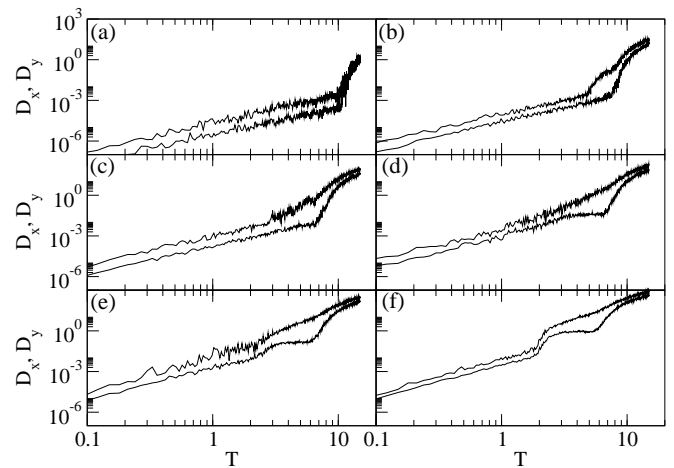}
\caption{
Diffusion in $x$ ($D_x$, bottom curves) 
and $y$ ($D_y$, top curves) directions for fillings
$B/B_\phi=$
(a) 1, (b) 4, (c) 7, (d) 10, (e) 15, and (f) 27.
}
\label{diffusexy}
\end{figure}

\begin{figure*}
\includegraphics[width=5.0in]{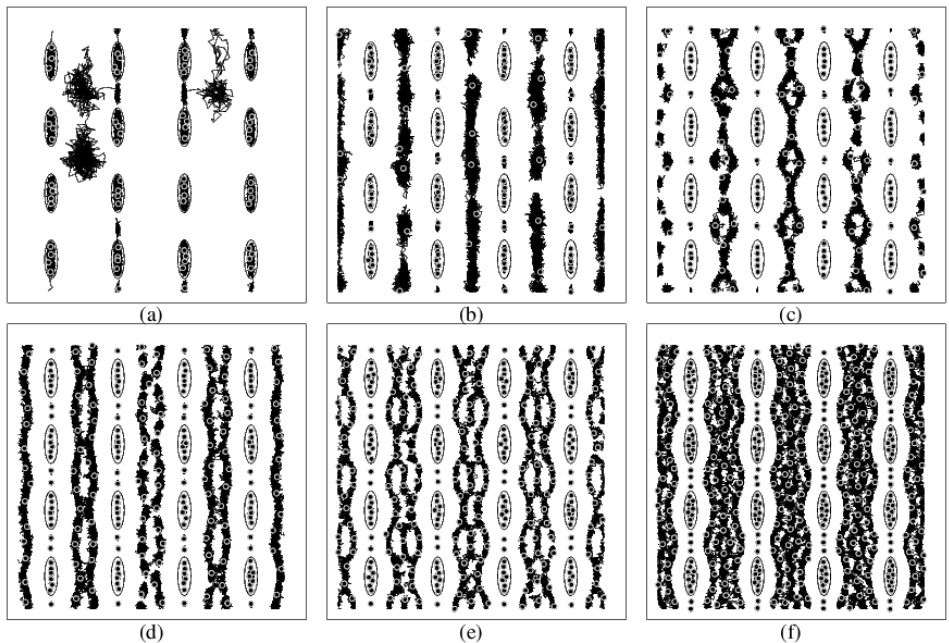}
\caption{
Illustration of vortex motion during the anisotropic melting process.
Ellipses indicate the locations of the pins, black dots represent
vortex positions, and lines show the trajectories of the vortices over
a fixed period of time.
(a) $B/B_\phi=4$, $T=7.05$.  (b) $B/B_\phi=7$, $T=3.90$.
(c) $B/B_\phi=10$, $T=2.46$.  (d) $B/B_\phi=15$, $T=2.34$.
(e) $B/B_\phi=20$, $T=1.86$.  (f) $B/B_\phi=27$, $T=2.25$.
}
\label{meltimage}
\end{figure*}

Depinning in the $y$ direction occurs at much lower applied driving
forces than depinning in the $x$ direction.  In addition, when we
zoom in on the
small driving force regime, shown 
in Fig.~\ref{fig:depiny}, we observe a nonmonotonic
dependence of $f_c^y$ on field that is caused by differences in the vortex
configurations.  
At fillings $B/B_\phi>7$,
the $y$ direction depinning forces are similar in scale to the vortex-vortex
interaction forces, permitting the system to be sensitive to vortex
configuration changes.  This is in contrast to the case for $x$
direction depinning where the extremely strong pinning forces dominate
the response of the system.
The $y$ direction depinning force $f_c^y$ is enhanced for
$B/B_\phi=10$, 20, and 27, as indicated in Fig.~\ref{fig:depiny}. 
To understand the origin of this enhancement,
we plot the vortex trajectories at driving forces just above
depinning for representative fields in Fig.~\ref{fig:fcypix}.
At $B/B_\phi=8$ in Fig.~\ref{fig:fcypix}(a), the vortices in the
interstitial channels form straight columns and can flow directly in the
$y$ direction upon depinning.  For $B/B_\phi=10$, the buckling of the
interstitial vortex channels
into the spaces between the pinning rows disrupts the straight
interstitial columns of vortices, and the vortex paths upon depinning
in the $y$ direction wind irregularly.  A higher driving force must be
applied to produce flow through winding channels compared to the
straight channels, since depinning can no longer occur via simple
propagation of a solitonlike vortex pulse.  

As the number of vortices inside the interstitial
channel increases, more than one channel of flow can open, as
illustrated in Fig.~\ref{fig:fcypix}(c) for $B/B_\phi=19$.  This 
multiple channel flow
occurs relatively easily and is associated with a low value of
$f_c^y$.  At $B/B_\phi=20$, where we observed 
well formed rows of vortices parallel
to the $x$ axis in Fig.~\ref{imagefig3}(b), the vortex flow is hindered
for depinning in the $y$ direction since the initial depinning paths
wind back and forth, as shown in
Fig.~\ref{fig:fcypix}(d).  Addition of one more vortex per unit cell 
distorts the rows of vortices parallel to the $x$ axis and permits
a smoother channel flow to occur with lower $f_c^y$, as illustrated
in Fig.~\ref{fig:fcypix}(e) for $B/B_\phi=21$.
Finally, we observe an increase in $f_c^y$ at the highest field we
studied, $B/B_\phi=27$.  In this case, the columns of vortex motion at
depinning do not mix; instead, the 
interstitial vortices move through five parallel channels, as shown
in Fig.~\ref{fig:fcypix}(f).  Two of the channels contain only a single
vortex per unit cell, and these two channels help to effectively jam
the flow at lower driving forces.

We next consider the melting properties of the vortex configurations
at zero applied driving force.
We measure the total diffusion 
$D_t=|{\bf r}(t_1)-{\bf r}(t_0)|/\delta t$
as a function of temperature by calculating
the distance traveled by the vortices during a fixed time period
$\delta t=t_1-t_0$.  For a fixed choice of $\delta_t$, melting appears
as a noticeable change in the slope of $D_t$ with temperature.  Figure
\ref{diffuse} shows representative $D_t$ vs $T$ curves for several
values of $B/B_\phi$.  The melting temperature decreases monotonically
with $B/B_\phi$.  

Due to the strong anisotropy of the pinning structure,
the melting occurs anisotropically.  To measure this, we compute
the diffusion in the $x$ and $y$ directions separately,
$D_x=|r_x(t_1)-r_x(t_0)|/\delta t$ and
$D_y=|r_y(t_1)-r_y(t_0)|/\delta t$.  We show representative plots
of $D_x$ and $D_y$ for six values of $B/B_\phi$ in Fig.~\ref{diffusexy}.
At $B/B_\phi=1$, in Fig.~\ref{diffusexy}(a), the vortices remain confined
within the pins for $T<10$, but since the pins are more extended in the
$y$ direction than in the $x$ direction, $D_y$ is larger than $D_x$ at
low temperatures, indicating that the vortices undergo larger thermal
excursions within the pins in the $y$ direction.
At $B/B_\phi=4$, shown in Fig.~\ref{diffusexy}(b), the 
change of slope in $D_y$
occurs at lower temperatures than the 
corresponding slope change in $D_x$.  In this case,
thermal activation causes a small number of the vortices to jump out of the
wells and occupy interstitial sites once $T \approx 6$.  These vortices
remain confined within a particular interstitial channel but are able to
diffuse slowly in the $y$ direction from one interstitial site to another, 
as illustrated in Fig.~\ref{meltimage}(a).
There is no significant $x$ direction diffusion until $T \approx 8$, 
when all of the vortices begin to hop in and out of the pinning sites.
When interstitial vortices are present from the beginning, 
such as at $B/B_\phi=7$ shown in Fig.~\ref{diffusexy}(c),  
$D_y$ is significantly larger than $D_x$ over a sizable
temperature range.
This is indicative of the formation of an 
interstitial liquid, where 
vortices diffuse freely in the $y$ direction in one-dimensional channels
between the columns of pins, but the $x$ direction vortex diffusion
is suppressed.
A representative example of the interstitial liquid
is plotted in Fig.~\ref{meltimage}(b)
for $B/B_\phi=7$.
A plateau begins
to form in $D_x$ for $B/B_\phi=10$ and above, as illustrated in
Fig.~\ref{diffusexy}(d).  At these higher fields, the interstitial channel 
contains more than one column of vortices, and the onset of $y$ direction 
diffusion within the interstitial channel is accompanied by a fixed amount of
$x$ translation due to the formation of ringlike structures in the 
vortex trajectories, as shown in Fig.~\ref{meltimage}(c).  
$D_x$ remains fixed at a constant value while
$D_y$ continues to increase with temperature as long as the 
interstitial liquid exists.  Similar behavior occurs for
two interstitial vortex channels at
$B/B_\phi=15$, as shown in Fig.~\ref{diffusexy}(e) and
Fig.~\ref{meltimage}(d), as well as 
for three interstitial vortex channels at $B/B_\phi=20$, plotted in
Fig.~\ref{meltimage}(e), and 
for four interstitial vortex channels at $B/B_\phi=27$,
illustrated in Fig.~\ref{diffusexy}(f) and Fig.~\ref{meltimage}(f).

\begin{figure}
\includegraphics[width=3.5in]{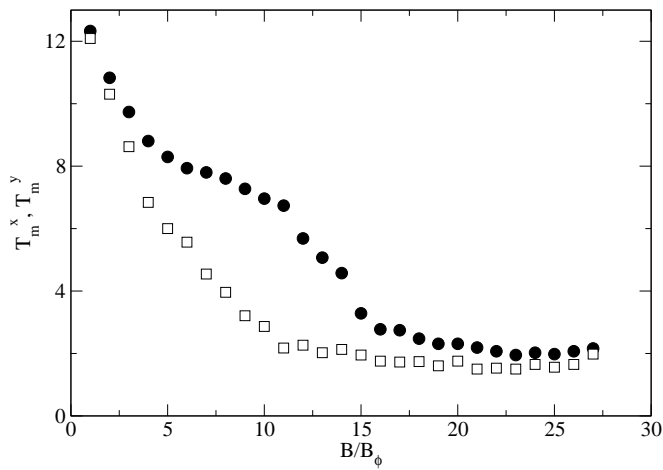}
\caption{
Squares: Melting temperature in the $y$ direction, $T_m^y$,
as measured from $D_y$.  Circles: Melting temperature in the $x$
direction, $T_m^x$, as measured from $D_x$.
}
\label{meltdiffuse}
\end{figure}

In Fig.~\ref{meltdiffuse} we summarize the behavior of the melting 
temperatures in the $x$ and $y$ directions, $T_m^x$ and $T_m^y$, 
as determined from the diffusion.
The melting occurs at high temperatures and is nearly isotropic for
$B/B_\phi=1$, but anisotropy develops rapidly as the field is increased
and becomes particularly pronounced once interstitial vortices
appear at $B/B_\phi=5$.  The $y$-direction melting temperature,
$T_m^y$, drops rapidly with field for $5 < B/B_\phi < 10$ as the 
interstitial vortex channel develops, while $T_m^x$ changes only
slowly with field in this regime.  Once the interstitial channels have
fully formed and begin to buckle into a configuration that is two or
more vortices wide, at $B/B_\phi \ge 10$, $T_m^y$ levels off and
$T_m^x$ begins to drop more rapidly as the space between interstitial
channel vortices and pinned vortices decreases with field.  Above
$B/B_\phi=15$, when the vortex configurations inside the pins
begin to buckle, $T_m^x$ also levels off at a low value.

The melting process at higher fillings in general proceeds as follows.
The first stage is contour melting, in which the particles diffuse along
the columns of pinning and maintain their spacing in columns.  Exchange
between columns happens at points where the columns are narrow, which
generally falls in between pinning sites.  As the temperature increases,
this channel structure is lost and the system forms a series of
liquid channels where all structure is destroyed.  The particles inside
the pins move freely inside the pins but remain confined.  The final
stage of melting occurs when the pinned particles depin and begin to move along
the pinning channel at the same time as particles begin to jump in the
$x$ direction from the pinning channel into the unpinned channel liquid and
vice versa.

\begin{figure}
\includegraphics[width=3.5in]{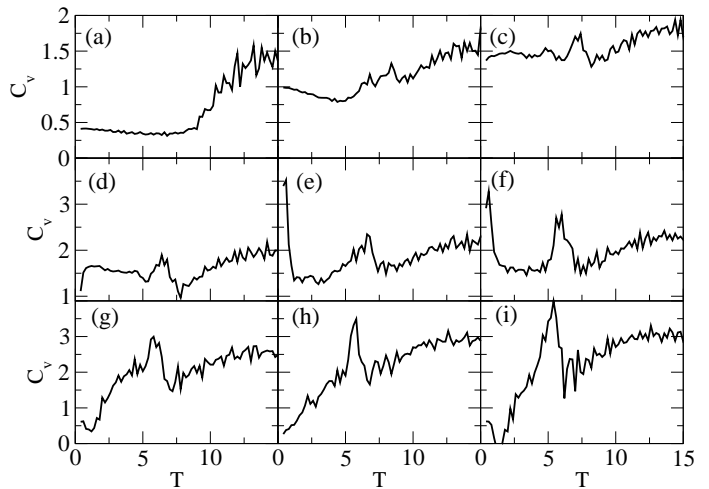}
\caption{
Specific heat $C_v$ for selected fillings
$B/B_\phi=$ (a) 2, (b) 5, (c) 9, (d) 11, (e) 14, (f) 16, (g) 20,
(h) 25, and (i) 27.
}
\label{cv1}
\end{figure}

The multiple stages of melting in the two directions also appear as
signatures in the specific heat,
\begin{equation} 
C_v(T)=(\langle E(T)^2\rangle - \langle E(T) \rangle^2)/T^2
\end{equation}
obtained from the average energy of the system $E(T)$.
To measure $C_v$, we begin with a $T=0$ vortex configuration obtained
from a slow anneal.  We then 
gradually increase the temperature in increments of 
$\Delta T$.  At each temperature we compute the average energy of the system, 
$E$, and average $E(T)$ over 20 to 600 realizations per filling fraction.  
We performed the largest
number of realizations for small values of $B/B_\phi$.  We
obtain the specific heat $C_v$ by taking the derivative of $E(T)$.
The resulting curves $C_v(T)$ are illustrated for selected fillings
in Fig.~\ref{cv1}.  For $B/B_\phi<5$, 
melting occurs when the vortices are thermally activated out of the
pinning sites.  This produces a single step up in $C_v$ at relatively
high temperatures near $T \sim 10$, as illustrated in Fig.~\ref{cv1}(a)
for $B/B_\phi=2$.  Once interstitial vortices appear,
as in Fig.~\ref{cv1}(b) for $B/B_\phi=5$, a new signature appears in 
$C_v$ at a lower temperature $T\sim 6$, corresponding to 
the temperature at which when the interstitial
vortices begin to diffuse and form an anisotropic interstitial liquid.
The magnitude of this second peak in $C_v$ varies with filling, as
shown in Fig.~\ref{cv1}(d) through (i), and the peak shifts to lower
temperature as $B/B_\phi$ increases, consistent with the decreasing
melting temperature shown in Fig.~\ref{meltdiffuse}.  Due to the
presence of the pinning, all of the melting transitions in this system
should be thermally activated crossovers.

In summary, we used numerical simulations to
study the configurations, dynamics, and melting properties of
vortex lattices interacting with elliptical pinning sites for both
low and high matching fields.  The configurations we obtain agree well
with experimental vortex images from the system we are modeling, and
as in the experiment we find that the saturation vortex number of the
individual pins increases with applied field.  The depinning thresholds
are highly anisotropic, particularly at higher fields, and the vortex
configurations upon depinning are very different depending on whether
the current is applied along or perpendicular to the pinning channels.
The depinning threshold in the $x$ direction 
occurs at large driving forces and decreases monotonically
with $B/B_\phi$, while the depinning threshold in the $y$ direction 
occurs at low driving forces and shows nonmonotonic behavior in 
response to the vortex configurations at each field.
We find anisotropic diffusion when the system melts, along with a
second structure in the specific heat corresponding to the presence
of an anisotropic interstitial vortex liquid at intermediate
temperatures.

We thank G. Karapetrov for useful discussions. 
This work was supported by the U.S.~Department of Energy under
Contract No.~W-7405-ENG-36.

\end{document}